\documentclass[doublecol]{epl2} 

\usepackage{amsfonts}
\usepackage{mathrsfs}
\usepackage{amsmath}
\usepackage{geometry}
\usepackage{graphicx}
\usepackage{verbatim}
\usepackage{appendix}

\title{Evolutionary dynamics of group cooperation with asymmetrical environmental feedback}
\shorttitle{Group cooperation with asymmetrical environmental feedback} 

\author{Yanxuan Shao\inst{1} \and Xin Wang\inst{2,3}\thanks{E-mail: \email{Xin.Wang-2@dartmouth.edu} }\and Feng Fu\inst{3,4}\thanks{E-mail: \email{feng.fu@dartmouth.edu}} }
\shortauthor{Y. Shao \etal}

\institute{                    
  \inst{1} School of Physics, Nankai University, Tianjin, 300071, China\\
  \inst{2} LMIB, NLSDE, BDBC and School of Mathematical Sciences, Beihang University, Beijing 100191, China\\
  \inst{3} Department of Mathematics, Dartmouth College, Hanover, NH 03755, USA\\
  \inst{4} Department of Biomedical Data Science, Geisel School of Medicine at Dartmouth, Lebanon, NH 03756, USA
}


\abstract{
In recent years, there has been growing interest in studying evolutionary games with environmental feedback. Previous studies exclusively focus on two-player games. However, extension to multi-player game is needed to study problems such as microbial cooperation and crowdsourcing collaborations. Here, we study coevolutionary public goods games where strategies coevolve with the multiplication factors of group cooperation. Asymmetry can arise in such environmental feedback, where games organized by focal cooperators may have a different efficiency than the ones by defectors. Our analysis shows that coevolutionary dynamics with asymmetrical environmental feedback can yield oscillatory convergence to persistent cooperation, if the relative changing speed of cooperators' multiplication factor is above a certain threshold. Our work provides useful insights into sustaining group cooperation in a changing world. }

\begin{document}

\maketitle


\section{Introduction}
Cooperation is a prominent phenomenon that widely exists in natural systems among various scales, ranging from microbes to human societies\cite{crespi2001evolution,west2003cooperation,riley2002bacteriocins,west2006social}. Evolutionary game theory is a powerful theoretical approach to study and understand why and on what conditions a persistent cooperation situation would occur\cite{smith1973logic,hofbauer1998evolutionary,nowak2006evolutionary,nowak2004emergence,danku2019knowing,du2011partner,wu2013adaptive}. In particular, public goods game (PGG) provides valuable insight into group cooperation\cite{boyd1988evolution,hauert1997effects,liu2018evolutionary,ginsberg2019evolution}, considering the great challenges we face nowadays, like global warming, pollution control and overexploitation of natural sources\cite{cohen1995population,hauser2014cooperating}. 

PGG can be seen as an extension of the Prisoner's Dilemma\cite{dawes1980social,kollock1998social}. In the classical PGG model with well-mixed interactions, the population eventually evolves into a mutual defection state where cooperation vanishes, under the assumption that individuals always choose rational strategies based on the incentives\cite{nash1950bargaining}. A great amount of works concentrate on solving this well-known tragedy of the commons\cite{hardin1968tragedy,feeny1990tragedy} by taking into account different realistic factors and evolving mechanisms in the game, such as kin selection\cite{hamilton1964genetical}, punishment and reward\cite{fehr2002altruistic,chen2015competition}, direct and indirect reciprocity theories\cite{trivers1971evolution,axelrod1981evolution,nowak1998dynamics,leimar2001evolution,fu2008reputation}, and in particular spatial reciprocity\cite{vukov2011incipient, szolnoki2010reward, wang2013interdependent, hauert2004spatial}. Additionally, a variety of factors describing heterogeneity of players have been incorporated, such as optional participation mechanism\cite{hauert2006evolutionary},  network topology structure\cite{gomez2011evolutionary,rong2009effect}, wealth-based selection\cite{chen2016individual} and different environment of the population\cite{hauert2019asymmetric}. These studies proved that heterogeneity plays an important role in promoting group cooperation in the real world\cite{santos2008social,wu2014social}.

Recently, a new framework of replicator dynamics with feedback-evolving games has been proposed to characterize the phenomenon that environment and individual behavior coevolves in many social-ecological and psychological-economic systems\cite{akcay2018collapse,cortez2018destabilizing,stewart2014collapse,chen2018punishment, hauert2006evolutionary, wakano2009spatial, gore2009snowdrift, perc2010coevolutionary}. The environmental feedback can result in oscillating dynamics of both environment quality and strategy states\cite{weitz2016oscillating}. The persistent cycles also occur in asymmetric conditions with a heterogeneous environment\cite{hauert2019asymmetric}. A more general framework for eco-evolutionary games shows that the cyclic dynamics only occurs under the condition that the environmental change is slow enough compared to strategy dynamics\cite{tilman2019evolutionary}. These models provide deep insights into the cooperation behavior in coevolutionary systems\cite{szolnoki2018environmental}. 

With the rapid development of network technology, crowdsourcing project, which is a new form of online collaboration aiming to complete a project by soliciting contributions from a large group of people or online communities, has attracted increasing attention in recent years\cite{zhang2016dynamics}. Crowdsourcing has been successfully applied into many fields, such as knowledge discovery and management, crisis mapping, and crowdfunding\cite{brabham2013crowdsourcing,meier2012crisis,belleflamme2014crowdfunding}. Interestingly, these online cooperations often happen under the preliminary conditions that there exists an authoritative organizer who leads the game and may decide the global payoff distributions to some extent, which is an important new character. In most crowdsourcing cases, cooperations are encouraged through a higher payoff structure for cooperators, such as extra incentives in commercial project or preferential access in knowledge discovery, which causes the emergence of asymmetrical feedback. We then raise an important question: on what conditions will these collaborations form successfully in general sense? Specifically, from the perspective of collaboration organizers, how to change the synergy effect of group of cooperators to encourage cooperation when the total resource and benefit of the project is restrained?

In this paper, we focus on the scenario where the multiplication factor of cooperators $r_c$ coevolves with the strategies in PGG (Similarly we also consider coevolving $r_d$ for defectors, see details in \emph{Appendix B}). We would like to see how ratios of total payoffs of cooperators vs defectors affect evolving adaptive environment. We let the multiplication factor of cooperators, $r_c$, changes in response to the global payoff difference between cooperators and defectors in the system, and in turn the multiplication factor, $r_c$, affects the evolutionary dynamics of individual cooperation behaviors. In this way, we add in the role of authoritative organizers who aim to organize the collaboration and can enforce the global payoff distributions, as described above. We highlight the conclusion that the feedback-evolving evolution can give rise to oscillating convergence to persistent cooperation in some parameter regime, but only if the relative changing speed of cooperators' multiplication factor exceeds a threshold. This result indicates that this asymmetrical environmental feedback in PGG is effective for group cooperation only when the feedback updates quickly and promptly enough compared to the strategy change. Our work sheds light on how to successfully organize a group collaboration and avoid the traps of social dilemma in projects like crowdsourcing.

\section{Model}

\begin{figure}[!h]
    \centering
    \includegraphics[width=0.45\textwidth]{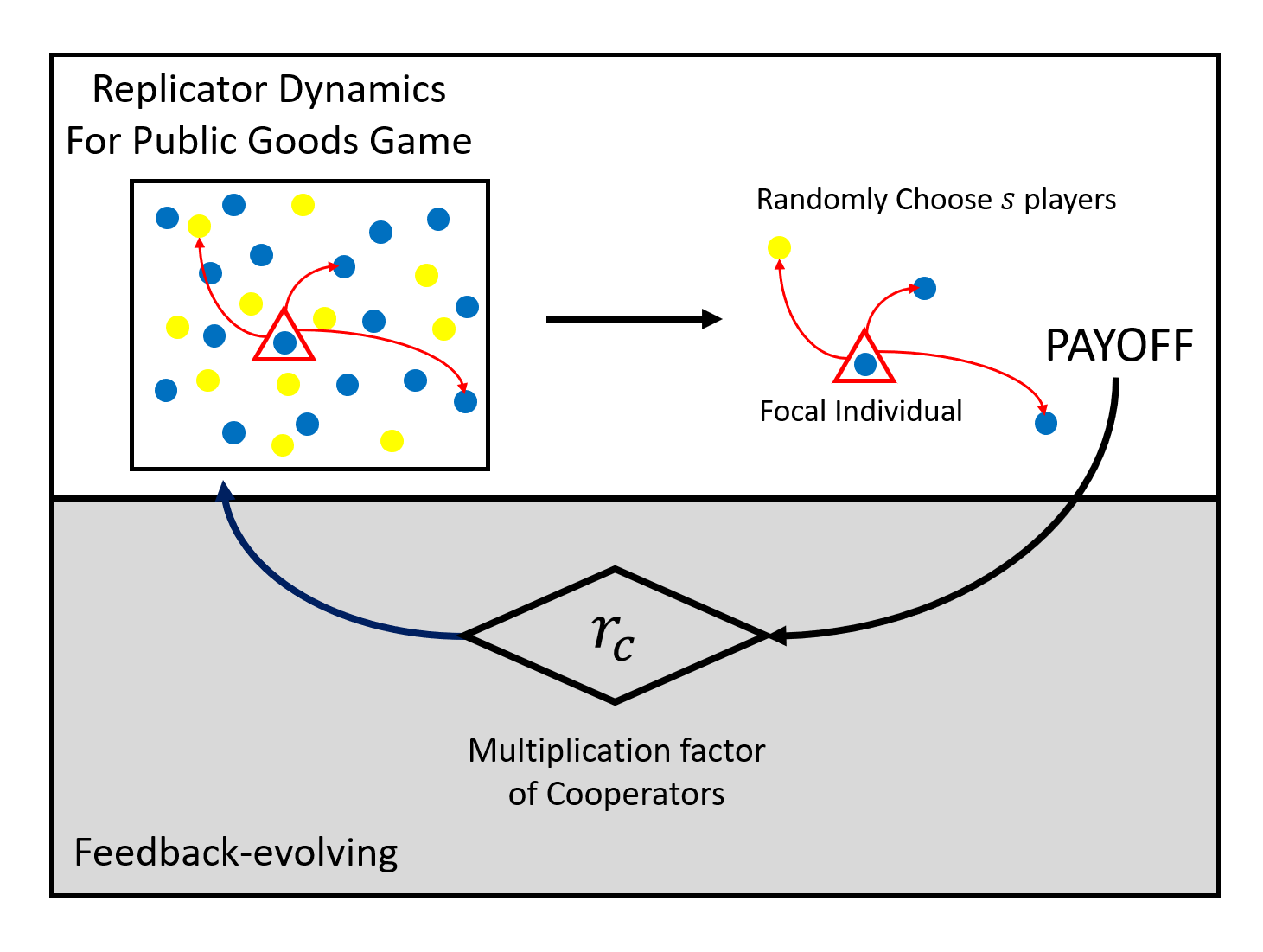}
    \caption{Schematic of the asymmetrical environmental feedback mechanism of our model. Graph on the top shows how the population evolves according to replicator dynamics, and meanwhile, total payoff of cooperators and defectors gives feedback to cooperator's multiplication factor and in turn modifies the dynamics of strategies.}
    \label{diagram}
\end{figure}

We consider PGG in a well-mixed infinitely large population, with each individual choosing to be a cooperator or defector, who contributes to the public pool or not, respectively. In each game, one focal individual randomly chooses other \(s\) players to play the game, which means there are in total \(s+1\) participants in one game. In classical PGG, the total contribution is multiplied by a multiplication factor $r$ and distributes to every participant equally. Here to characterize the asymmetrical environmental feedback, we assume defector's multiplication factor $r_d$ keeps constant, which is relatively low, and cooperator's multiplication factor $r_c$ keeps changing depending on the influence of global payoff distribution. In turn, the changing $r_c$ affects individual's payoff and drives dynamics of strategies, as illustrated in fig. \ref{diagram}. 

We denote \(x\) as the frequency of cooperators in the population. For simplicity and without loss of generality, we let the contribution of each cooperator be 1. For a focal individual, the possibility that \(m\) out of \(s\) selected individuals are cooperators is
\begin{equation}
	\binom{s}{m}x^m(1-x)^{s-m}
\end{equation}
Thus expectation of the focal individual's payoff is 
\begin{equation}
    \begin{split}
        P_c &= \sum\limits_{m=0}^{s}\binom{s}{m}x^m(1-x)^{s-m}\left[\frac{(m+1)r_c}{s+1}-1\right] \\
        &= -1 + \frac{(1+sx)}{s+1}r_c 
\end{split}
\end{equation}
\begin{equation}
	\begin{split}
        P_d &= \sum\limits_{m=0}^{s}\binom{s}{m}x^m(1-x)^{s-m}\frac{mr_d}{s+1}\\
        &= \frac{sx}{s+1}r_d
    \end{split}
\end{equation}

Replicator dynamics are widely used in evolutionary games, which describes the time evolution of the frequency of each strategy. Here the replicator equation for $x$, the frequency of cooperators in the population, is
\begin{equation}
	\begin{split}
		\dot{x} &= x(P_c-\bar{P})\\
		&= x(1-x)\left(\frac{sx+1}{s+1}r_c-\frac{sx}{s+1}r_d-1\right)
	\end{split}
\end{equation}
where \(\bar{P}\) denotes the average payoff of the population. 

The other equation describing dynamics of cooperator's multiplication factor is 
\begin{equation}
	\dot{r_c} = \epsilon (r_c-\alpha)(\beta-r_c) f(x,r_c)
\end{equation}
where \(f(x,r_c)\) describes the feedback mechanism of total payoff in game interaction with its sign deciding whether \(r_c\) increases or decreases. \(\alpha\) and \(\beta\) denote minimum and maximum values of multiplication factor of cooperators, therefore by the term \((r_c-\alpha)(\beta-r_c)\), \(r_c\) will grow logistically and be confined to the range \([\alpha,\beta]\). According to social dilemma of PGG, we have \(1<\alpha<\beta<s+1\). \(\epsilon\) denotes the relative changing speed of \(r_c\) compared to \(x\). The multiplication factor of cooperators, which is characterized by \(r_c\), in turn influences the payoffs as well as the frequencies of different strategies, resulting in a feedback loop. We assume the multiplication factor of cooperators is modified by global payoffs due to the limitation of total rewards for the project and the zero-sum characteristic of resource consumption:
\begin{equation}
	f(x,r_c) = -xP_c+\theta(1-x)P_d
\end{equation}
where \(xP_c\) and \((1-x)P_d\) are the global payoff for cooperator and defector in the population, respectively. \(\theta>0\) denotes the ratio of increasement rate to decreasment rate of cooperator's and defector's total payoff expectation in the system. Here, when resource is adequate, cooperators are rewarded according to their contributions to the public pool, while depletion of resources in the crowdsourcing prevent cooperator's multiplication factor increasing infinitely. 

Thus the ODE systems for our model can be written as:
\begin{equation}
	\begin{cases}
	\begin{split}
		\dot{x} &= x(1-x)\left(\frac{sx+1}{s+1}r_c-\frac{sx}{s+1}r_d-1\right)\\
		\dot{r_c} &= \epsilon(r_c-\alpha)(\beta-r_c)\left[-x\left(-1+\frac{r_c(1+sx)}{s+1}\right)\right.\\
		&\left.+\theta(1-x)\frac{r_dsx}{s+1}\right]
	\end{split}
	\end{cases}
	\label{differential_equation}
\end{equation}

\section{Results}

\textit{Stability of fixed points and thresholds of multiplication factors}

There are six possible fixed points of the model: five are on the boundary and the remaining one is an interior fixed point. For the five boundary fixed points, only two of them can be stable: (i) (\(x^*=0\)), the population is dominated by defectors and is always stable, which also occurs in the classic model of PGG; and (ii) (\(x^*=1, r_c=\alpha\)), which is stable only if the multiplication factor of defectors, $r_d$, is smaller than the threshold $r_d^* = \frac{(s+1)(\alpha-1)}{s}$. This possible fixed point corresponds to the state where the population is dominated by cooperators and the multiplication factor of cooperators $r_c$ is at its minimum value. 

\begin{figure}[!h]
    \centering
    \includegraphics[width=0.49\textwidth]{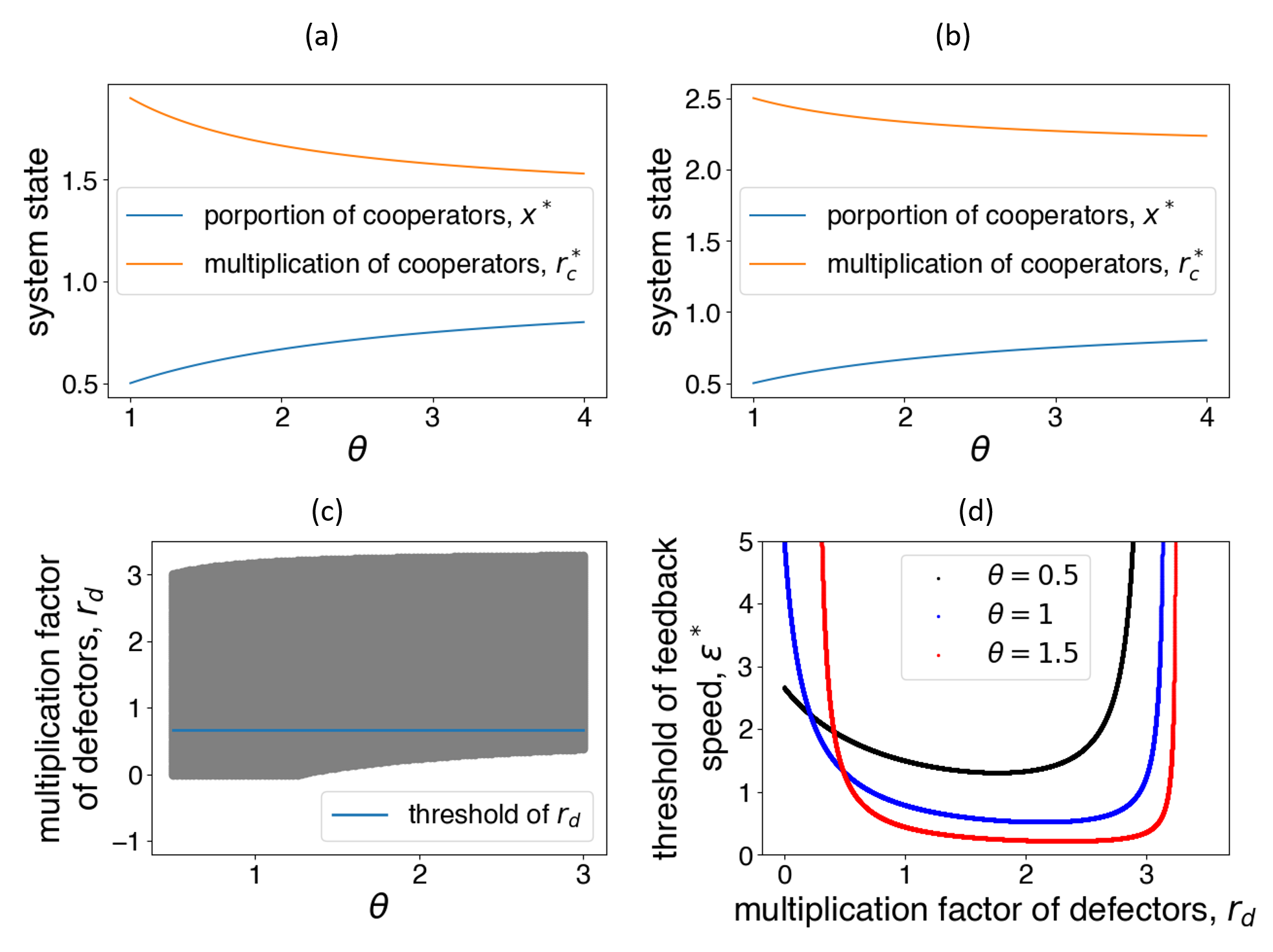}
    \caption{Impact of varying model parameters on population equilibrium states. Panel (a) and (b) show changes of the interior fixed point (\(x^*\) and \(r_c^*\)) as \(\theta\) increases, with \(r_d=0.5\) for (a) and \(r_d=1.5\) for (b). (c) presents the critical boundary of $r_d$ as well as the threshold $r_d^*$ as $\theta$ increases. (d) shows the dependence of the threshold of relative changing speed \(\epsilon^*\) on  \(\theta\) and \(r_d\). In all panels, $s=3, \alpha=1.5, \beta=3.5$. }
    \label{threshold}
\end{figure}

Besides, there is one interior fixed point which can be stable: 
\begin{equation}
	\begin{cases}
	\begin{split}
		x^*&=\frac{\theta}{1+\theta}\\
		r_c^*&=\frac{\theta r_ds+(s+1)(\theta+1)}{\theta s+\theta+1}
	\end{split}
	\end{cases}
	\label{interior_fixed_point}
\end{equation}
It corresponds to a stable population composed by both cooperators and defectors, with a medium value of cooperators' multiplication factor.  Eq. \ref{interior_fixed_point} indicates that the final position of interior fixed point is only influenced by $\theta$, the ratio of increasement rate to decreasement rate of cooperator’s and defector’s global payoff, which characterizes the nature of the project itself. The detailed impacts of $\theta$ on \(x^*\) and \(r_c^*\) are shown in fig. \ref{threshold}(a)(b). We fix $s=3, \alpha=1.5, \beta=3.5$ and we have $r_d^*=\frac{2}{3}$. Therefore we set \(r_d=0.5\) for  fig. \ref{threshold}(a) and \(r_d=1.5\) for  fig. \ref{threshold}(b), respectively. Results show that when \(\theta\) increases, the stable frequency of cooperators \(x^*\) increases, while the cooperator's multiplication factor \(r_c^*\) decreases. In order to get an intuitive understanding, we raise an example of a team work. If the team work does not require strong abilities of the workers, like pure labour work, \(\theta\) increases accordingly, calling for more people participating in the team work for better outcome. Eventually, there will be a higher proportion of cooperators with a relatively low multiplication factor of cooperators. On the contrary, if team members are expected to be more skilled, like in scientific collaborations, the decrease of \(\theta\) asks for people who can make real contributions to the project, resulting in lower frequency of cooperators with higher multiplication factor of cooperators.
 
Since \(\alpha\leq r_c\leq \beta\), the interior fixed point is meaningful only when 
\begin{equation*}
	\max\{\frac{\alpha(\theta s+\theta+1)-(s+1)(\theta+1)}{\theta s},0\} \leq r_d
\end{equation*}
\begin{equation}
	\leq \frac{\beta(\theta s+\theta+1)-(s+1)(\theta+1)}{\theta s}
	\label{rd}
\end{equation}
In fig. \ref{threshold}(c), we show critical boundary of $r_d$ as well as the position of threshold $r_d^*$ as $\theta$ increases.

Finally, this interior fixed point is stable only when \(\epsilon>\epsilon^*\), in which $ \epsilon^*$ depends on other parameters \(s, r_d, \theta, \alpha\) and \(\beta\), which writes 
\begin{equation}
	\epsilon^*=\frac{(1-x^*)s(r_c^*-r_d)}{(sx^*+1)(r_c^*-\alpha)(\beta-r_c^*)}
	\label{epsilon_thredshold}
\end{equation} 
The interior fixed point is the center of limit cycle (\(\epsilon=\epsilon^*\)) or is unstable (\(\epsilon<\epsilon^*\)) otherwise. In fig. \ref{threshold}(d), we set $s=3, \alpha=1.5, \beta=3.5$ and show how $\epsilon^*$ varies as $\theta$ and $r_d$ change. We choose three values for \(\theta\): \(0.5,1\) and \(1.5\), in which condition \(0\leq r_d\leq 3\), \(0 \leq r_d\leq \frac{19}{6}\) and \(\frac{1}{9}\leq r_d \leq \frac{29}{9}\), respectively. $\epsilon^*$ firstly goes down sharply followed by mild decreases and a steep increase, causing by the logistic term \((r_c-\alpha)(\beta-r_c)\). In realistic games we concern more about the situations where the multiplication factor of defectors is neither too large nor too small and \(\epsilon^*\) does not change much. Besides, \(\epsilon^*\) is larger when \(\theta\) is smaller.  

The detailed proof for the stability of all six fixed points using Jacobian matrices is shown in \emph{Appendix A}.\\

\textit{Detailed conditions for the emergence of persistent cooperation}

\begin{figure*}[!h]
    \centering
    \includegraphics[width=\textwidth]{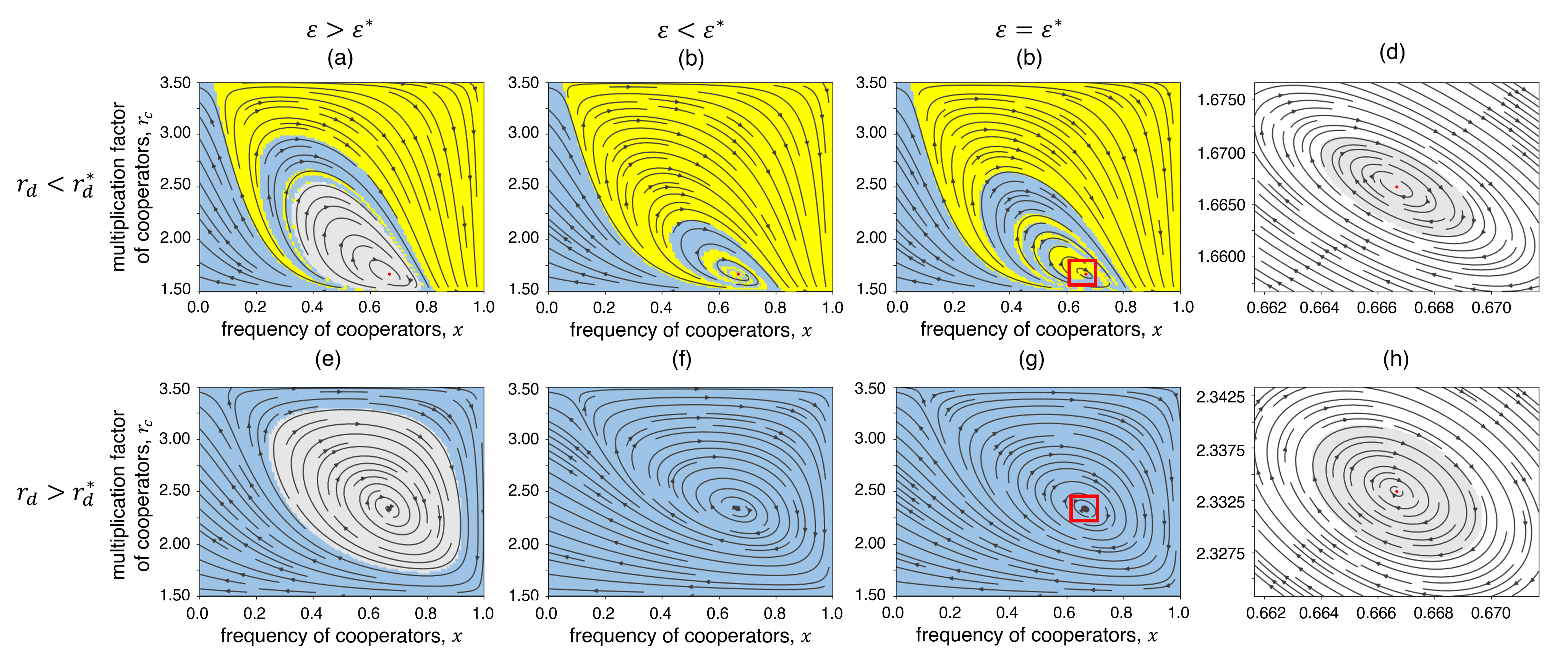}
    \caption{Phase plane dynamics of \(x-r_c\) system with \(s=3, \alpha=1.5, \beta=3.5, \theta=2\). For the first row, \(r_d=0.5\), \(\epsilon=1.5, \frac{14}{11},1\). For the second row, \(r_d=1.5\), \(\epsilon=0.4,\frac{2}{9},0.2\). Graphs in the last column are magnification of the area near interior fixed points, highlighted with red rectangles in the third column. Blue, yellow and grey areas show attracting fields of different fixed points (\(x^*=0\)), (\(x^*=1,r_c^*=\alpha\)) and (\(x^*=\frac{\theta}{\theta+1},r_c^*=\frac{\theta r_ds+(s+1)(\theta+1)}{\theta s+\theta+1}\)), respectively. }
    \label{phase_graph}
\end{figure*}

In fig. \ref{phase_graph}, we show how asymmetrical environmental feedback mechanism affects system state under different situations using phase graphs. We choose a group of parameters $s=3, \alpha=1.5, \beta=3.5, \theta=2$ and we have $r_d^*=\frac{2}{3}$ accordingly. Therefore we set \(r_d=0.5,1.5\) for two rows separately, where the thresholds of \(\epsilon\) are \(\frac{14}{11}\) and \(\frac{2}{9}\) correspondingly, according to Eq. \ref{epsilon_thredshold}. Fig. \ref{phase_graph}(a)-(c) show that a mutual cooperation state can always occur as long as $r_d<r_d^*$, which means when defector's multiplication factor is much lower than cooperator's, the asymmetrical environmental feedback mechanism can effectively promote the emergence of group collaboration. However, this condition can rarely be satisfied in real world, especially in PGG where one can hardly control defector's payoffs. Therefore, we concern more about the emergence of persistent co-existence of cooperators and defectors, i.e., the stability condition of interior fixed point. The comparison of fig. \ref{phase_graph}(a) vs (b)(c) as well as (e) vs (f)(g) reveals that the relative changing speed of cooperator's multiplication factor \(\epsilon\) determines the stability of interior fixed point. The persistent co-existence of cooperators and defectors can only emerge when \(\epsilon\) exceeds a threshold \(\epsilon^*\). In particular, \(\epsilon>\epsilon^*\) is the only chance for breaking social dilemma when $r_d>r_d^*$, as shown in fig. \ref{phase_graph}(e)-(g). Therefore, we conclude that the asymmetrical environmental feedback in PGG is effective for the emergence of group cooperation only when the feedback updates quickly enough compared to the strategy dynamics.

\begin{figure}[!h]
    \centering
    \includegraphics[width=0.49\textwidth]{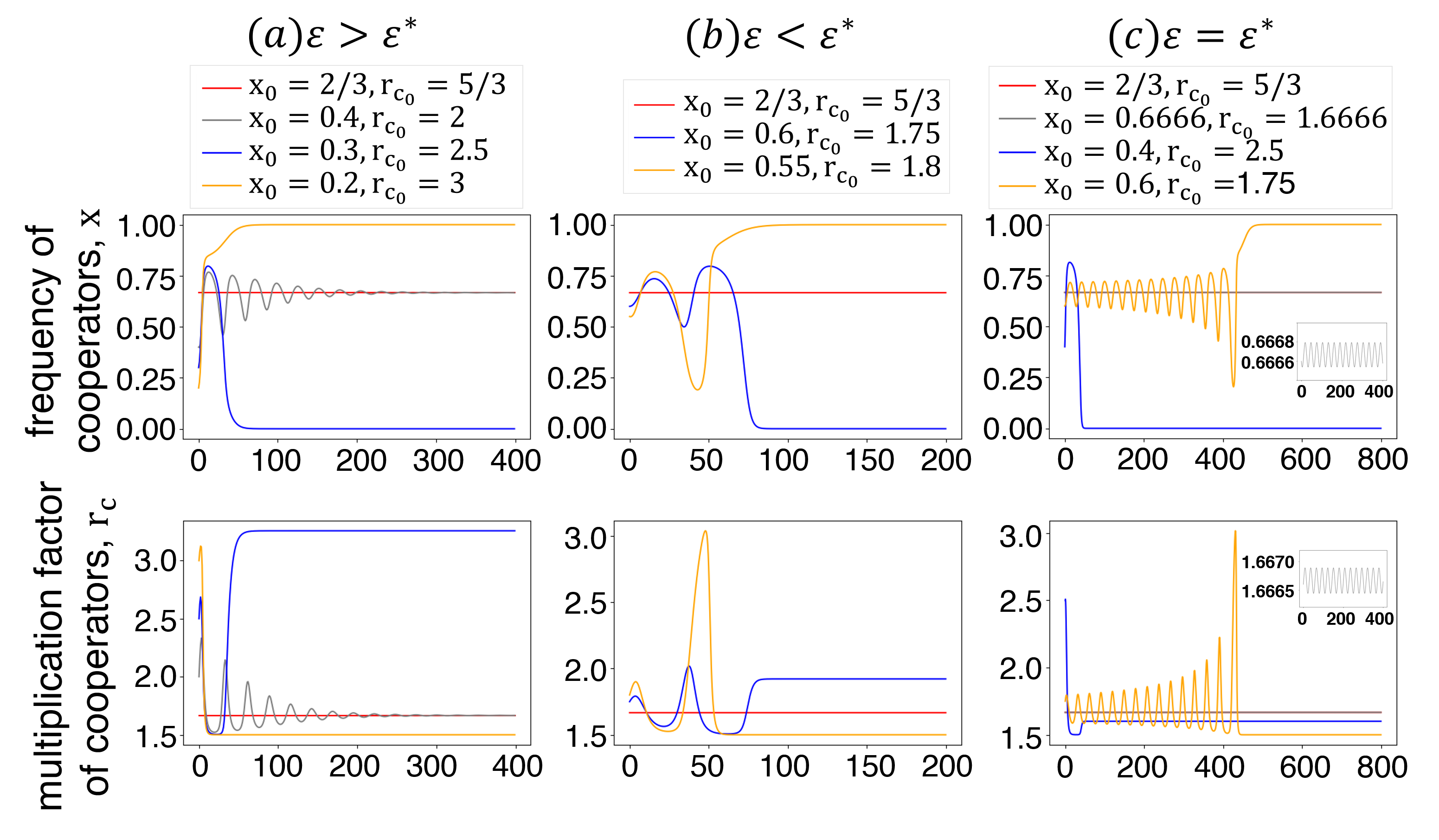}
    \caption{Time evolutions of the system state, representing by the frequency of strategies and the multiplication factor of cooperators, under different initial conditions. Parameters are \(s=3,\alpha=1.5,\beta=3.5,\theta=2,r_d=0.5\) and \(\epsilon=1.5,\frac{14}{11},1\) for (a)-(c), respectively. Initial conditions are shown in each graph. Inserted small graphs in the third column are magnification of the dynamics of initial point \(x_0=0.6666, r_{c_0}=1.6666\).}
    \label{dynamics}
\end{figure}

Furthermore, in fig. \ref{dynamics}, we present time evolutions of different system states, representing by the frequency of strategies and the multiplication factor of cooperators, corresponding to the situations shown in fig. \ref{phase_graph}(a)-(c). Parameters are \(s=3,\alpha=1.5,\beta=3.5,\theta=2,r_d=0.5\) and \(\epsilon=1.5,\frac{14}{11},1\) for fig. \ref{dynamics}(a)-(c) respectively. Here $r_d<r_d^*$. In fig. \ref{dynamics}(a), \(\epsilon>\epsilon^*\), which means the asymmetrical environmental feedback is quickly and promptly enough, the interior fixed point is stable. A population with initial conditions near the interior fixed point experiences oscillating convergence to the interior equilibrium state. Other initial states far from the interior fixed point experience rapid oscillation and converge to the boundary, either cooperation-dominated or defector-dominated. In fig. \ref{dynamics}(b), when \(\epsilon<\epsilon^*\), which indicates the cooperators are not rewarded in time and all initial states oscillate to the boundary. In fig. \ref{dynamics}(c), interior fixed point becomes center of the limit cycle when \(\epsilon\) is exactly at its threshold, in which situation a tiny range of initial state of the system experiences regular and continuous oscillation around the center. Initial point from the rest part of the domain ends either in defector-dominated or in cooperator-dominated population. 

\section{Conclusion}
In this paper, we extend the two-player evolutionary games with environmental feedback to multi-player situation where strategies coevolve with the multiplication factor of group cooperation. Using coevolutionary PGG framework, we study a new form of collaboration in real world. To describe the existence of asymmetry in the games where focal organizers who aim to organize the collaboration may enforce the global payoff distributions, we differentiate the multiplication factor of cooperators and defectors. On one hand, cooperators are encouraged in order to avoid the social dilemma by increasing their multiplication factor. While on the other hand, resource will limit the number of cooperators when their global payoff is large enough. By adding this asymmetrical environmental feedback mechanism to replicator dynamics, the population can oscillatorily converge to a cooperator-defector coexisting state if the relative changing speed of cooperator's multiplication factor exceeds a threshold, breaking tragedy of the commons in traditional PGG. The final frequency of cooperators is determined only by one parameter: the ratio of increasement rate to decreasement rate of cooperator's and defector's global payoff, involving with the limitation of total resource and the zero-sum characteristic of resource consumption. If the ratio is relatively large, which means cooperation in the project is not that resource-consuming, it appears to be more cooperators in the population and they get a relatively low multiplication factor. Conversely, there are fewer cooperators who get higher multiplication factor if the ratio is small, corresponding to projects which are resource-consuming. 

This asymmetrical environmental feedback mechanism well describes collaboration situations like crowdsourcing projects, which has potential applications to explain a number of real-world cooperation phenomena. Our work also shows detailed conditions for the emergence of stable cooperation with resource restraints, thereby shedding light on organizing a successful group collaboration under similar circumstances. While current results focus on linear PGG, a potential direction for further studies is extending our framework to nonlinear PGG, such as threshold PGG which has been successfully used to better understand human behaviors in response to the climate change\cite{zhu2017evolutionary,wang2009emergence,zhang2013tale}.

\acknowledgments
X.W. gratefully acknowledges generous support by the China Scholarship Council. F.F. gratefully acknowledges the G. Norman Albree Trust Fund, the Dartmouth Faculty Startup Fund, the Neukom CompX Faculty Grant, Walter \& Constance Burke Research Initiation Award and NIH Roybal Center Pilot Grant.


\begin{thebibliography}{10}
\expandafter\ifx\csname url\endcsname\relax\def\url#1{\texttt{#1}}\fi

\bibitem{crespi2001evolution}
\Name{Crespi B.~J.} \REVIEW{Trends in ecology \& evolution}{16}{2001}{178}.

\bibitem{west2003cooperation}
\Name{West S.~A. \and Buckling A.} \REVIEW{Proceedings of the Royal Society of
  London. Series B: Biological Sciences}{270}{2003}{37}.

\bibitem{riley2002bacteriocins}
\Name{Riley M.~A. \and Wertz J.~E.} \REVIEW{Annual Reviews in
  Microbiology}{56}{2002}{117}.

\bibitem{west2006social}
\Name{West S.~A., Griffin A.~S., Gardner A. \and Diggle S.~P.} \REVIEW{Nature
  reviews microbiology}{4}{2006}{597}.

\bibitem{smith1973logic}
\Name{Smith J.~M. \and Price G.~R.} \REVIEW{Nature}{246}{1973}{15}.

\bibitem{hofbauer1998evolutionary}
\Name{Hofbauer J. \and Sigmund K.} \Book{Evolutionary games and population
  dynamics} (Cambridge university press) 1998.

\bibitem{nowak2006evolutionary}
\Name{Nowak M.~A.} \Book{Evolutionary dynamics} (Harvard University Press)
  2006.

\bibitem{nowak2004emergence}
\Name{Nowak M.~A., Sasaki A., Taylor C. \and Fudenberg D.}
  \REVIEW{Nature}{428}{2004}{646}.

\bibitem{danku2019knowing}
\Name{Danku Z., Perc M. \and Szolnoki A.} \REVIEW{Scientific
  reports}{9}{2019}{262}.

\bibitem{du2011partner}
\Name{Du F. \and Fu F.} \REVIEW{Dynamic Games and Applications}{1}{2011}{354}.

\bibitem{wu2013adaptive}
\Name{Wu T., Fu F., Zhang Y. \and Wang L.} \REVIEW{Scientific
  reports}{3}{2013}{1550}.

\bibitem{boyd1988evolution}
\Name{Boyd R. \and Richerson P.~J.} \REVIEW{Journal of theoretical
  Biology}{132}{1988}{337}.

\bibitem{hauert1997effects}
\Name{Hauert C. \and Schuster H.~G.} \REVIEW{Proceedings of the Royal Society
  of London. Series B: Biological Sciences}{264}{1997}{513}.

\bibitem{liu2018evolutionary}
\Name{Liu L., Wang S., Chen X. \and Perc M.} \REVIEW{Chaos: An
  Interdisciplinary Journal of Nonlinear Science}{28}{2018}{103105}.

\bibitem{ginsberg2019evolution}
\Name{Ginsberg A. \and Fu F.} \REVIEW{Games}{10}{2019}{1}.

\bibitem{cohen1995population}
\Name{Cohen J.~E.} \REVIEW{Science}{269}{1995}{341}.

\bibitem{hauser2014cooperating}
\Name{Hauser O.~P., Rand D.~G., Peysakhovich A. \and Nowak M.~A.}
  \REVIEW{Nature}{511}{2014}{220}.

\bibitem{dawes1980social}
\Name{Dawes R.~M.} \REVIEW{Annual review of psychology}{31}{1980}{169}.

\bibitem{kollock1998social}
\Name{Kollock P.} \REVIEW{Annual review of sociology}{24}{1998}{183}.

\bibitem{nash1950bargaining}
\Name{Nash~Jr J.~F.} \REVIEW{Econometrica: Journal of the Econometric
  Society}{}{1950}{155}.

\bibitem{hardin1968tragedy}
\Name{Hardin G.} \REVIEW{science}{162}{1968}{1243}.

\bibitem{feeny1990tragedy}
\Name{Feeny D., Berkes F., McCay B.~J. \and Acheson J.~M.} \REVIEW{Human
  ecology}{18}{1990}{1}.

\bibitem{hamilton1964genetical}
\Name{Hamilton W.~D.} \REVIEW{Journal of theoretical biology}{7}{1964}{17}.

\bibitem{fehr2002altruistic}
\Name{Fehr E. \and G{\"a}chter S.} \REVIEW{Nature}{415}{2002}{137}.

\bibitem{chen2015competition}
\Name{Chen X., Szolnoki A. \and Perc M.} \REVIEW{Physical Review
  E}{92}{2015}{012819}.

\bibitem{trivers1971evolution}
\Name{Trivers R.~L.} \REVIEW{The Quarterly review of biology}{46}{1971}{35}.

\bibitem{axelrod1981evolution}
\Name{Axelrod R. \and Hamilton W.~D.} \REVIEW{science}{211}{1981}{1390}.

\bibitem{nowak1998dynamics}
\Name{Nowak M.~A. \and Sigmund K.} \REVIEW{Journal of theoretical
  Biology}{194}{1998}{561}.

\bibitem{leimar2001evolution}
\Name{Leimar O. \and Hammerstein P.} \REVIEW{Proceedings of the Royal Society
  of London. Series B: Biological Sciences}{268}{2001}{745}.

\bibitem{fu2008reputation}
\Name{Fu F., Hauert C., Nowak M.~A. \and Wang L.} \REVIEW{Physical Review
  E}{78}{2008}{026117}.

\bibitem{vukov2011incipient}
\Name{Vukov J., Santos F.~C. \and Pacheco J.~M.} \REVIEW{PLoS
  One}{6}{2011}{e17939}.

\bibitem{szolnoki2010reward}
\Name{Szolnoki A. \and Perc M.} \REVIEW{EPL (Europhysics
  Letters)}{92}{2010}{38003}.

\bibitem{wang2013interdependent}
\Name{Wang Z., Szolnoki A. \and Perc M.} \REVIEW{Scientific
  reports}{3}{2013}{1183}.

\bibitem{hauert2004spatial}
\Name{Hauert C. \and Doebeli M.} \REVIEW{Nature}{428}{2004}{643}.

\bibitem{hauert2006evolutionary}
\Name{Hauert C., Holmes M. \and Doebeli M.} \REVIEW{Proceedings of the Royal
  Society B: Biological Sciences}{273}{2006}{2565}.

\bibitem{gomez2011evolutionary}
\Name{Gomez-Gardenes J., Romance M., Criado R., Vilone D. \and S{\'a}nchez A.}
  \REVIEW{Chaos: An Interdisciplinary Journal of Nonlinear
  Science}{21}{2011}{016113}.

\bibitem{rong2009effect}
\Name{Rong Z. \and Wu Z.-X.} \REVIEW{EPL (Europhysics
  Letters)}{87}{2009}{30001}.

\bibitem{chen2016individual}
\Name{Chen X. \and Szolnoki A.} \REVIEW{Scientific reports}{6}{2016}{32802}.

\bibitem{hauert2019asymmetric}
\Name{Hauert C., Saade C. \and McAvoy A.} \REVIEW{Journal of theoretical
  biology}{462}{2019}{347}.

\bibitem{santos2008social}
\Name{Santos F.~C., Santos M.~D. \and Pacheco J.~M.}
  \REVIEW{Nature}{454}{2008}{213}.

\bibitem{wu2014social}
\Name{Wu T., Fu F., Dou P. \and Wang L.} \REVIEW{Physica A: Statistical
  Mechanics and its Applications}{413}{2014}{86}.

\bibitem{akcay2018collapse}
\Name{Akcay E.} \REVIEW{Nature communications}{9}{2018}{2692}.

\bibitem{cortez2018destabilizing}
\Name{Cortez M., Patel S. \and Schreiber S.} \REVIEW{bioRxiv}{}{2018}{488759}.

\bibitem{stewart2014collapse}
\Name{Stewart A.~J. \and Plotkin J.~B.} \REVIEW{Proceedings of the National
  Academy of Sciences}{111}{2014}{17558}.

\bibitem{chen2018punishment}
\Name{Chen X. \and Szolnoki A.} \REVIEW{PLoS computational
  biology}{14}{2018}{e1006347}.

\bibitem{wakano2009spatial}
\Name{Wakano J.~Y., Nowak M.~A. \and Hauert C.} \REVIEW{Proceedings of the
  National Academy of Sciences}{106}{2009}{7910}.

\bibitem{gore2009snowdrift}
\Name{Gore J., Youk H. \and Van~Oudenaarden A.}
  \REVIEW{Nature}{459}{2009}{253}.

\bibitem{perc2010coevolutionary}
\Name{Perc M. \and Szolnoki A.} \REVIEW{BioSystems}{99}{2010}{109}.

\bibitem{weitz2016oscillating}
\Name{Weitz J.~S., Eksin C., Paarporn K., Brown S.~P. \and Ratcliff W.~C.}
  \REVIEW{Proceedings of the National Academy of Sciences}{113}{2016}{E7518}.

\bibitem{tilman2019evolutionary}
\Name{Tilman A.~R., Plotkin J. \and Akcay E.} \REVIEW{bioRxiv}{}{2019}{493023}.

\bibitem{szolnoki2018environmental}
\Name{Szolnoki A. \and Chen X.} \REVIEW{EPL (Europhysics
  Letters)}{120}{2018}{58001}.

\bibitem{zhang2016dynamics}
\Name{Zhang Z.-K., Liu C., Zhan X.-X., Lu X., Zhang C.-X. \and Zhang Y.-C.}
  \REVIEW{Physics Reports}{651}{2016}{1}.

\bibitem{brabham2013crowdsourcing}
\Name{Brabham D.~C.} \Book{Crowdsourcing} (Mit Press) 2013.

\bibitem{meier2012crisis}
\Name{Meier P.} \REVIEW{Journal of Map \& Geography Libraries}{8}{2012}{89}.

\bibitem{belleflamme2014crowdfunding}
\Name{Belleflamme P., Lambert T. \and Schwienbacher A.} \REVIEW{Journal of
  business venturing}{29}{2014}{585}.

\bibitem{zhu2017evolutionary}
\Name{Zhu Y., Zhang J., Sun Q. \and Chen Z.} \REVIEW{Knowledge-Based
  Systems}{130}{2017}{51}.

\bibitem{wang2009emergence}
\Name{Wang J., Fu F., Wu T. \and Wang L.} \REVIEW{Physical Review
  E}{80}{2009}{016101}.

\bibitem{zhang2013tale}
\Name{Zhang Y., Fu F., Wu T., Xie G. \and Wang L.} \REVIEW{Scientific
  reports}{3}{2013}{2021}.

\end{thebibliography}

\clearpage 

\end{document}



\section{Appendix A: Stability of fixed points in asymmetrical environmental feedback model}

Differential equations describing the whole system are
\begin{equation}
    \begin{cases}
        \begin{split}
            \dot{x} &= x(1-x)\left(\frac{sx+1}{s+1}r_c-\frac{sx}{s+1}r_d-1\right)\\
            \dot{r_c} &= \epsilon(r_c-\alpha)(\beta-r_c)\left[-x\left(-1+\frac{r_c(1+sx)}{s+1}\right)+\theta(1-x)\frac{r_dsx}{s+1}\right] 
        \end{split}
    \end{cases}
\end{equation}

Jacobian of this system is
\footnotesize{}
\begin{equation*}
	\left[\begin{matrix}
    (1-2x)\left(\frac{sx+1}{s+1}r_c-\frac{sx}{s+1}r_d-1\right)+x(1-x)\frac{s(r_c-r_d)}{s+1} & x(1-x)\frac{sx+1}{s+1}\\
    \epsilon(r_c-\alpha)(\beta-r_c)\left[1-\frac{r_c(1+2sx)}{s+1}+\theta\frac{r_ds(1-2x)}{s+1}\right] & \epsilon x\left[(\alpha+\beta-2r_c)\left(1-\frac{r_c(1+sx)}{s+1}+\theta(1-x)\frac{r_ds}{s+1}\right)-(r_c-\alpha)(\beta-r_c)\frac{(1+sx)}{s+1}\right]
	\end{matrix}\right]
\end{equation*}

\normalsize{}
Letting the derivatives of \(x\) and \(r_c\) to be 0 and considering the preconditions, we can find several fixed points of this system.

(1) \(x^*=0\)
\begin{equation*}
    J(x^*=0) = \left[\begin{matrix}
    \frac{r_c}{s+1}-1 & 0\\
    \epsilon(r_c-\alpha)(\beta-r_c)\left(1-\frac{r_c}{s+1}+\theta\frac{r_ds}{s+1}\right) & 0
    \end{matrix}\right]
\end{equation*}

Eigenvalues are \(\lambda_1=\frac{r_c-(s+1)}{s+1}<0\) (since \(r_c\leq\beta<s+1\)) and \(\lambda_2=0\). Therefore the fixed point is stable.

(2) \(x^*=1,r_c^*=\alpha\)
\begin{equation*}
    J(1,\alpha) = \left[\begin{matrix}
    \frac{r_ds}{s+1}-\alpha+1 & 0\\
    0 & \epsilon(\beta-\alpha)(1-\alpha)
    \end{matrix}\right]
\end{equation*}

Eigenvalues are \(\lambda_1=\frac{r_ds}{s+1}-\alpha+1\) and \(\lambda_2=\epsilon(\beta-\alpha)(1-\alpha)<0\). This fixed point is stable when 
\begin{equation}
	r_d<\frac{(s+1)(\alpha-1)}{s}
\end{equation}

(3) \(x^*=1,r_c^*=\beta\)
\begin{equation*}
    J(1,\beta) = \left[\begin{matrix}
    \frac{r_ds}{s+1}-\beta+1 & 0\\
    0 & \epsilon(\alpha-\beta)(1-\beta)
    \end{matrix}\right]
\end{equation*}

Eigenvalues are \(\lambda_1=\frac{r_ds}{s+1}-\beta+1\) and \(\lambda_2=\epsilon(\beta-\alpha)(\beta-1)>0\). This fixed point is unstable. 

(4) \(x^*=\frac{s+1-\alpha}{s(\alpha-r_d)},r_c^*=\alpha\)
\begin{equation*}
    J(x^*,\alpha) = \left[\begin{matrix}
    x^*(1-x^*)\frac{s(\alpha-r_d)}{s+1} & x^*(1-x^*)\frac{sx^*+1}{s+1}\\
    0 & \epsilon(\beta-\alpha)x^*\left[1-\frac{\alpha(1+sx^*)}{s+1}+\theta(1-x^*)\frac{r_ds}{s+1}\right]
    \end{matrix}\right]
\end{equation*}

Eigenvalues are \(\lambda_1=x^*(1-x^*)\frac{s(\alpha-r_d)}{s+1}\) and \(\lambda_2=\epsilon(\beta-\alpha)x^*\left[1-\frac{\alpha(1+sx^*)}{s+1}+\theta(1-x^*)\frac{r_ds}{s+1}\right]\). When \(\lambda_1<0\), \(\alpha<r_d\), so \(\lambda_2=\epsilon(\beta-\alpha)\frac{\theta}{\theta+1}\frac{1}{s+1}\left[(s+1-\alpha)+\frac{s\theta}{\theta+1}(r_d-\alpha)\right]>0\). Therefore the two eigenvalues cannot be negative simultaneously and the fixed point is unstable. 

(5) \(x^*=\frac{s+1-\beta}{s(\beta-r_d)},r_c^*=\beta\)
\begin{equation*}
    J(x^*,\alpha) = \left[\begin{matrix}
    x^*(1-x^*)\frac{s(\beta-r_d)}{s+1} & x^*(1-x^*)\frac{sx^*+1}{s+1}\\
    0 & \epsilon(\alpha-\beta)x^*\left[1-\frac{\beta(1+sx^*)}{s+1}+\theta(1-x^*)\frac{r_ds}{s+1}\right]
    \end{matrix}\right]
\end{equation*}

Eigenvalues are \(\lambda_1=x^*(1-x^*)\frac{s(\beta-r_d)}{s+1}>0\) and \(\lambda_2=\epsilon(\alpha-\beta)x^*\left[1-\frac{\beta(1+sx^*)}{s+1}+\theta(1-x^*)\frac{r_ds}{s+1}\right]\). This fixed point is unstable. 

(6) Interior fixed point 
\begin{equation}
	\begin{cases}
    \begin{split}
        x^*&=\frac{\theta}{1+\theta}\\
        r_c^*&=\frac{\theta r_ds+(s+1)(\theta+1)}{\theta s+\theta+1}
    \end{split}
    \end{cases}
\end{equation}
\begin{equation*}
    J(x^*,r_c^*) = \left[\begin{matrix}
    x^*(1-x^*)\frac{s(r_c^*-r_d)}{s+1} & x^*(1-x^*)\frac{sx^*+1}{s+1}\\
    \epsilon(r_c^*-\alpha)(\beta-r_c^*)\left[1-\frac{r_c^*(1+2sx^*)}{s+1}+\theta\frac{r_ds(1-2x^*)}{s+1}\right] & -\epsilon x^*(r_c^*-\alpha)(\beta-r_c^*)\frac{sx^*+1}{s+1}
    \end{matrix}\right]
\end{equation*}

If we use 
\begin{equation*}\left[\begin{matrix}
	A&B\\C&D
\end{matrix}\right]\end{equation*}
to represent the complicated matrix mentioned above. There are two conditions for the interior fixed point to be stable: 
\begin{equation}
	\begin{cases}
	AD-BC&>0\\
	A+D&<0
	\end{cases}
\end{equation}

The first condition is always satisfied. Therefore, the interior fixed point is stable when 
\begin{equation}
\epsilon>\epsilon^*
\end{equation}
where
\begin{equation}
\begin{split}	
	\epsilon^*&=\frac{(1-x^*)s(r_c^*-r_d)}{(sx^*+1)(r_c^*-\alpha)(\beta-r_c^*)}
\end{split}
\end{equation}

To sum up, fixed point on the boundary \(x^*=0\) is always stable. Another fixed point on the boundary \(x^*=1,r_c^*=\alpha\) is stable when \(r_d<\frac{(s+1)(\alpha-1)}{s}\). Interior fixed point \(x^*=\frac{\theta}{\theta+1}, r_c^*=\frac{\theta r_ds+(s+1)(\theta+1)}{\theta s+\theta+1}\) is stable when \(\epsilon>\frac{(1-x^*)s(r_c^*-r_d)}{(sx^*+1)(r_c^*-\alpha)(\beta-r_c^*)}\). 
\\

\section{Appendix B: Population, cooperator's multiplication factor, and defector's multiplication factor coevolving model} 
In the main text, the multiplication factor of defectors is fixed. Here we discuss a more general model with coevolution of strategy dynamics and multiplication factor of both cooperators and defectors. Similar to the cooperator’s multiplication factor feedback model we have analyzed that cooperation behavior is encouraged when the proportion of cooperators is low while the amount of cooperators is limited due to the limitation of resources, the full model can be described using differential equations as:
\begin{equation}
	\begin{cases}
		\begin{split}
			\dot{x} &= x(1-x)(P_c-P_d)=F(x,r_c,r_d)\\
			\dot{r_c} &= \epsilon_1(r_c-\alpha)(\beta-r_c)\left[-xP_c+\theta(1-x)P_d\right]=G(x,r_c,r_d)\\
			\dot{r_d} &= \epsilon_2(r_d-\gamma)(\eta-r_d)\left[-\theta(1-x)P_d+xP_c\right]=H(x,r_c,r_d)
		\end{split}
	\end{cases}
\end{equation}
where \(\alpha\) (\(\gamma\)) and \(\beta\) (\(\eta\)) denote the minimum and maximum values of multiplication factor of cooperators (defectors). \(\epsilon_1\) and \(\epsilon_2\) denote relative changing speed of \(r_c\) and \(r_d\) compared to the ratio of cooperators in population. Input the forms of \(P_c\) and \(P_d\), equations turn
\begin{equation}
	\begin{cases}
		\begin{split}
			\dot{x} &= x(1-x)\left(\frac{sx+1}{s+1}r_c-\frac{sx}{s+1}r_d-1\right)\\
			\dot{r_c} &= \epsilon_1(r_c-\alpha)(\beta-r_c)\left[\theta(1-x)\frac{sxr_d}{s+1}-x\left(\frac{sx+1}{s+1}r_c-1\right)\right]\\
			\dot{r_d} &= \epsilon_2(r_d-\gamma)(\eta-r_d)\left[x\left(\frac{sx+1}{s+1}r_c-1\right)-\theta(1-x)\frac{sxr_d}{s+1}\right]
		\end{split}
	\end{cases}
\end{equation}
Therefore we have Jacobian for the system
\begin{equation*}
	J=\left[\begin{matrix}
		\frac{\partial{F}}{\partial{x}} & \frac{\partial{F}}{\partial{r_c}} & \frac{\partial{F}}{\partial{r_d}}\\
		\\
		\frac{\partial{G}}{\partial{x}} & \frac{\partial{G}}{\partial{r_c}} & \frac{\partial{G}}{\partial{r_d}}\\
		\\
		\frac{\partial{H}}{\partial{x}} & \frac{\partial{H}}{\partial{r_c}} & \frac{\partial{H}}{\partial{r_d}}
	\end{matrix}\right]
\end{equation*}

where
\begin{equation}
	\begin{split}
		\frac{\partial{F}}{\partial{x}} &= (1-2x)\left(\frac{sx+1}{s+1}r_c-\frac{sx}{s+1}r_d-1\right)+x(1-x)\frac{s(r_c-r_d)}{s+1}\\
		\frac{\partial{F}}{\partial{r_c}} &= x(1-x)\frac{sx+1}{s+1}\\
		\frac{\partial{F}}{\partial{r_d}} &= -x(1-x)\frac{sx}{s+1}\\
		\frac{\partial{G}}{\partial{x}} &= \epsilon_1(r_c-\alpha)(\beta-r_c)\left[1-\frac{1+2sx}{s+1}r_c+\theta\frac{(1-2x)s}{s+1}r_d\right]\\
		\frac{\partial{G}}{\partial{r_c}} &= \epsilon_1 x\left[(\alpha+\beta-2r_c)\left(1-\frac{r_c(1+sx)}{s+1}+\theta(1-x)\frac{r_ds}{s+1}\right)-(r_c-\alpha)(\beta-r_c)\frac{(1+sx)}{s+1}\right]\\
		\frac{\partial{G}}{\partial{r_d}} &= \epsilon_1(r_c-\alpha)(\beta-r_c)\theta(1-x)\frac{sx}{s+1}\\
		\frac{\partial{H}}{\partial{x}} &= \epsilon_2(r_d-\gamma)(\eta-r_d)\left[\frac{2sx+1}{s+1}r_c-1-\theta\frac{(1-2x)s}{s+1}r_d\right]\\
		\frac{\partial{H}}{\partial{r_c}} &= \epsilon_2(r_d-\gamma)(\eta-r_d)x\frac{sx+1}{s+1}\\
		\frac{\partial{H}}{\partial{r_d}} &= \epsilon_2(\gamma+\eta-2r_d)\left[x\left(\frac{sx+1}{s+1}r_c-1\right)-\theta(1-x)\frac{sx}{s+1}r_d\right]+ \epsilon_2(r_d-\gamma)(\eta-r_d)\left[-\theta(1-x)\frac{sx}{s+1}\right]
	\end{split}
\end{equation}

Letting derivatives of \(x,r_c\) and \(r_d\) to be 0, here are the conditions for fixed points. 

(1) \(x^*=0\)
\begin{equation*}
    J(x^*=0) = \left[\begin{matrix}
    \frac{r_c^*}{s+1}-1 & 0 & 0\\
    \epsilon_1(r_c^*-\alpha)(\beta-r_c^*)\left(1-\frac{\alpha}{s+1}+\theta\frac{r_d^*s}{s+1}\right) & 0 & 0\\
    \epsilon_2(r_d^*-\gamma)(\eta-r_d^*)\left(\frac{\alpha}{s+1}-1-\theta\frac{r_d^*s}{s+1}\right) & 0 & 0
    \end{matrix}\right]
\end{equation*}

Three eigenvalues are \(\lambda_1=\frac{r_c^*-(s+1)}{s+1}<0\) (since \(r_c^*\leq\beta<s+1\)) and \(\lambda_2=\lambda_3=0\). Therefore the fixed point is stable.

(2) \(x^*=1,r_c^*=\alpha,r_d^*=\gamma\)
\begin{equation*}
    J(1,\alpha,\gamma) = \left[\begin{matrix}
    \frac{s\gamma}{s+1}+1-\alpha & 0 & 0\\
    0 & \epsilon_1(\beta-\alpha)(1-\alpha) & 0\\
    0 & 0 & \epsilon_2(\eta-\gamma)(\alpha-1)
    \end{matrix}\right]
\end{equation*}

Three eigenvalues are \(\lambda_1=\frac{s\gamma}{s+1}+1-\alpha\), \(\lambda_2=\epsilon_1(\beta-\alpha)(1-\alpha)\), and \(\lambda_3=\epsilon_2(\eta-\gamma)(\alpha-1)>0\) (since \(1<\alpha<\beta<s+1, \gamma<\eta\)). Therefore the fixed point is unstable. 

(3) \(x^*=1,r_c^*=\alpha,r_d^*=\eta\)
\begin{equation*}
    J(1,\alpha,\eta) = \left[\begin{matrix}
    \frac{s\eta}{s+1}+1-\alpha & 0 & 0\\
    0 & \epsilon_1(\beta-\alpha)(1-\alpha) & 0\\
    0 & 0 & \epsilon_2(\gamma-\eta)(\alpha-1)
    \end{matrix}\right]
\end{equation*}

Eigenvalues are \(\lambda_1=\frac{s\eta}{s+1}+1-\alpha\), \(\lambda_2=\epsilon_1(\beta-\alpha)(1-\alpha)<0\), and \(\lambda_3=\epsilon_2(\gamma-\eta)(\alpha-1)<0\). The fixed point is stable when \(\eta<\eta^*\), with 
\begin{equation}
	\eta^*=\frac{(s+1)(\alpha-1)}{s}
\end{equation}

(4) \(x^*=1,r_c^*=\beta,r_d^*=\gamma/\eta\)

Three eigenvalues are \(\lambda_1=\frac{r_d^*s}{s+1}+1-\beta\), \(\lambda_2=\epsilon_1(\alpha-\beta)(1-\alpha)>0\) (since \(1<\alpha<\beta<s+1\)), and \(\lambda_3=\epsilon_2(\gamma+\eta-2r_d^*)(\beta-1)\). Therefore the fixed points are unstable. 
\small{}

(5) \(r_c^*=\alpha/\beta,r_d^*=\gamma/\eta,x^*=\frac{s+1-r_c^*}{s(r_c^*-r_d^*)}\)

If \(x^*\) exists, \(r_c^*>r_d^*\) always exists. One of the eigenvalue is \(\lambda_1=x^*(1-x^*)\frac{s(r_c^*-r_d^*)}{s+1}>0\). Therefore the fixed points are unstable. 

(6) Interior fixed point 
\begin{equation}
	\begin{cases}
			x^* =\frac{\theta}{\theta+1}\\
			\alpha \leq r_c^*\leq \beta\\
			\frac{\theta s\gamma+(\theta+1)(s+1)}{\theta s+\theta+1}\leq r_c^* \leq \frac{\theta s\eta+(\theta+1)(s+1)}{\theta s+\theta+1}\\
			r_d^* = \frac{(\theta s+\theta+1)r_c^*-(\theta+1)(s+1)}{\theta s}
	\end{cases}
\end{equation}
is stable when 
\begin{equation}
	\left(\frac{\partial{F}}{\partial{x}}+\frac{\partial{G}}{\partial{r_c}}+\frac{\partial{H}}{\partial{r_d}}\right)\bigg|_{x=x^*,r_c=r_c^*,r_d=r_d^*}<0
\end{equation}
that is 
\begin{equation}
	\frac{(\theta+1)(s+1)-(\theta+1)r_c^*}{\theta}<\epsilon_1(r_c^*-\alpha)(\beta-r_c^*)(\theta s+\theta+1) + \epsilon_2(r_d^*-\gamma)(\eta-r_d^*)\theta s
	\label{stable_condition}
\end{equation}

\begin{figure}[!h]
    \centering
    \includegraphics[width=0.95\textwidth]{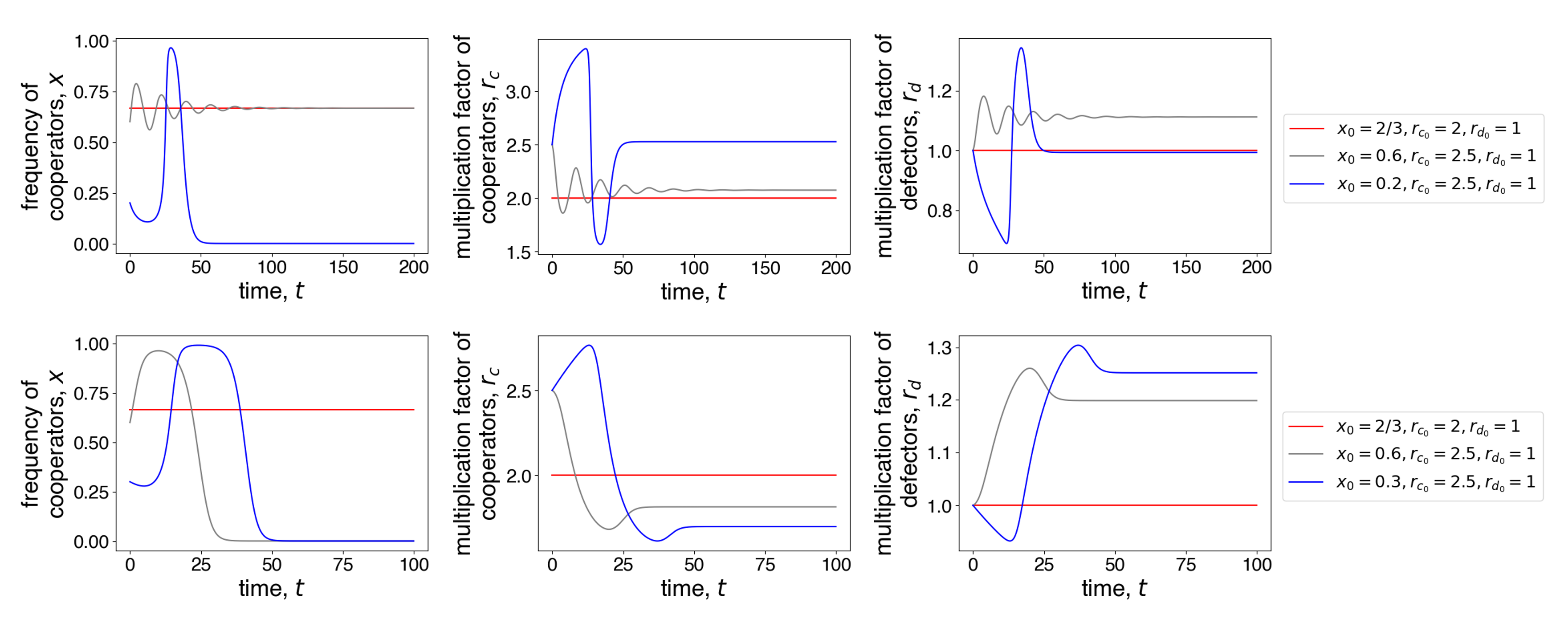}
    \caption{Evolutionary dynamics in population, cooperator's multiplication factor and defector's multiplication factor coevolving games with different initial conditions converging to corresponding fixed points. Parameters are \(s=3,\alpha=1.5,\beta=3.5,\gamma=0.5,\eta=1.5,\theta=2\). For the first row: \(\epsilon_1=\epsilon_2=0.5\); for the second row: \(\epsilon_1=\epsilon_2=0.1\). Different initial conditions are shown each graph.}
    \label{dynamics}
\end{figure}

In order to have a more explicit understanding of the system, we choose parameters $s=3, \alpha=1.5, \beta=3.5, \gamma=0.5, \eta=1.5, \theta=2$. The stable condition for interior fixed point turns 
\begin{equation}
	4-r_c^*<6\epsilon_1(r_c^*-1.5)(3.5-r_c^*) + 9\epsilon_2 \left(r_c^*-\frac{5}{3}\right)\left(\frac{7}{3}-r_c^*\right)
\end{equation}
and the value of \(r_c^*\) has a range of \(\left[\frac{5}{3},\frac{7}{3}\right]\). Population dynamics are demonstrated in fig. \ref{dynamics} with different \(\epsilon_1, \epsilon_2\) and initial conditions chosen. 

For the first row in fig. \ref{dynamics}, \(\epsilon_1=\epsilon_2=0.5\) and the stable condition for interior fixed point (eq. \ref{stable_condition}) turns \(30{r_c^*}^2-136r_c^*+149<0\), which is satisfied when \(r_c^*>\frac{34}{15}-\frac{\sqrt{154}}{30}\approx 1.853\). Therefore the interior fixed point is stable for most $r_c$ values here. At this situation, the speed of feedback from payoff is fast enough. System initialized from some parameter regimes experiences oscillation and finally evolves to the state with stable frequency of both cooperators and defectors. Here the final frequency of cooperators is only determined by parameter \(\theta\), which is influenced by the games themselves. Lager \(\theta\) is, lower ratio of cooperators will emerge in the population. Meanwhile, the stable multiplication factors of cooperators and defectors satisfy the following conditions and depend on specific initial conditions
\begin{equation}
	r_d^* =\frac{3r_c^*-4}{2}
\end{equation}
System initialized from other conditions ends in defector dominated state.  

While for the second row in fig. \ref{dynamics}, \(\epsilon_1=\epsilon_2=0.1\) and the stable condition for interior fixed point (eq. \ref{stable_condition}) turns \(30{r_c^*}^2-152r_c^*+213<0\), which is never satisfied. So the interior fixed point is unstable any time and the relative changing speed of multiplication factors of cooperators and defectors compared to population dynamics is not large enough. System initialized from conditions except the interior fixed point ends in state without cooperators, corresponding to results from classical PGG. 

In conclusion, here we observe similar phenomena to the coevolutionary model shown in the main text where the multiplication factor of defectors is fixed. The stable cooperating population emerges only if the relative feedback speed of multiplication factor is fast enough compared to evolution dynamics. And the final proportion of cooperators is only determined by \(\theta\), which is affected by the intrinsic game properties.